\documentclass[12pt,oneside]{JHEP3}
\usepackage{graphics}
\usepackage{amsfonts}
\usepackage[latin1]{inputenc}

\title{Geometric Transition as a Change of Polarization} 
\author{Sergio Monta\~nez\\ {\it  Instituto de F\'\i 
sica Te\'orica CSIC/UAM,}\\ {\it C-XVI Universidad Aut\'onoma,}\\ {\it E-28049 Madrid \rm 
Spain}\\
E-mail: 
\email{sergio.montannez@uam.es} }

\abstract{Taking the results of hep-th/0702110 we study the Dijkgraaf-Vafa open/closed topological string duality by considering the wavefunction behavior of the partition function. We find that the geometric transition associated with the duality can be seen as a change of polarization.
}

\preprint{
IFT-UAM/CSIC-07/26\\
{\tt arXiv:0705.2980}
}

\keywords{Topological Strings, Matrix Models}

\newcommand{\ba}{\begin{array}} 
\newcommand{\ea}{\end{array}} 
\newcommand{\be}{\begin{equation}} 
\newcommand{\ee}{\end{equation}}
\newcommand{\ben}{\begin{equation}} 
\newcommand{\een}{\end{equation}} 
\newcommand{\bea}{\begin{eqnarray}} 
\newcommand{\eea}{\end{eqnarray}} 
\newcommand{\gsim}{\mathrel{\mathop{\kern 0pt \rlap 
  {\raise.2ex\hbox{$>$}}} \lower.9ex\hbox{\kern-.190em $\sim$}}}

\def\entero{\mathbb{Z}}
\def\real{\mathbb{R}}
\def\complejo{\mathbb{C}}
\def\imaginario{\mathbb{I}}

\newcommand{\trace}{\,\mathrm{Tr}\,} 
\def\I{\mbox{Im}}
\def\R{\mbox{Re}}
\def\pd{\partial} 
\def\a{\alpha} 
\def\bet{\beta} 
\def\b{\beta}
\def\g{\gamma} 
\def\d{\delta}

\def\t{\tau} 
\def\l{\lambda}

\newcommand{\Ce}{\ensuremath{{\cal{C}}}}

\newcommand{\De}{\ensuremath{{\cal{D}}}}

\newcommand{\ra}{\rangle}
\newcommand{\la}{\langle}
\makeindex
\begin{document}

\section{Introduction}

Topological strings, which were introduced by Witten \cite{Witten:1988xj,Witten:1989ig}
 more that fifteen years ago, have led, not only to very interesting mathematical results, but also to important physical applications beyond those that originally motivated their construction. In addition, they can be considered as ``toy models'' helping us to understand some basic properties of physical string theory.

In fact, it was Witten \cite{Witten:1993ed} himself who, trying to face up the problem of background-dependence in string theory, found a very interesting result: the background-dependent partition function of closed B-model topological strings can be seen as a background-dependent representation of a background-independent state in a quantum mechanical system whose phase space is $H^3(M,\real)$, being $M$ the Calabi-Yau threefold target space.

Another important lesson we have learned about topological strings is that there is a large $N$ topological string duality \cite{Gopakumar:1998ki,Ooguri:2002gx}, associated with a kind of geometric transitions, relating different open and closed string backgrounds. To be more precise, let us consider the well-known proposal of Dijkgraaf and Vafa \cite{Dijkgraaf:2002fc}. The starting point is the resolved local CY threefold $M_{\rm res}$ encoded by the complex curve
\bea
y^2 - W^{\prime}(x)^2=0 &  & (x,y) \in \complejo^2
\eea
where $W^\prime (x)= \prod_{a=1}^{d}(x-x_a)$ is a polynomial of degree $d$. The authors consider that there are $N_a$ B-model branes wrapping the $\complejo {P}^1_a$ obtained after blowing up the $x=x_a$ singularity. In this case, the open string field theory governing the dynamics of the open topological strings attached to the branes reduces to the holomorphic matrix model
\be
\label{mamo}
Z_{\rm MM}(g_s,N)=\frac{1}{\rm{vol}(U(N))} \int \De M \exp \left[-\frac{1}{g_s} \trace W(M) \right]
\ee
where $N=\sum_{a=1}^{d} N_a$ and $g_s$ is the topological string coupling. More precisely, the open topological string partition function corresponds to the perturbative 't Hooft expansion of this matrix model around a vacuum at which there are $N_a$ eigenvalues surrounding the critical point $x_a$,
\be
Z_{\rm open}(g_s,N_a)= \exp \left(\sum_{g=0} g_s^{2g-2} \sum_{h_1,...,h_n=1} F_{g,h_1,h_2,...,h_n} t_1^{h_1}t_2^{h_2}...t_n^{h_n}\right)
\ee
where $t_a=g_sN_a$ are the 't Hooft couplings. Dijkgraaf and Vafa conjectured that the 't Hooft resummation of the free energies
\be
F^{\rm open}_g(t)=\sum_{h_1,...,h_n=1} F_{g,h_1,h_2,...,h_n} t_1^{h_1}t_2^{h_2}...t_n^{h_n}
\ee  
computes the closed topological string free energies $F^{\rm closed}_g(t)$ on the background $M_{\rm{def}}$, the deformed CY associated with the classical spectral curve of the matrix model. That is,
\be
\label{DVconjecture}
F^{\rm open}_g(t)=F^{\rm closed}_g(t)
\ee
where the quantities $t^a$s in the closed side are identified with the complex structure deformation parameters. This conjecture has been tested in refs. \cite{Klemm:2002pa,Dijkgraaf:2002yn,Huang:2006si}.

At this point the first naive problem comes by noticing that the $F^{\rm open}_g(t)$ are naturally holomorphic functions, whereas $F^{\rm closed}_g(t,\bar{t})$ have a non-holomorphic dependence given by the holomorphic anomaly \cite{Bershadsky:1993ta,Bershadsky:1993cx}. Therefore, a natural question is what happens in eq. (\ref{DVconjecture}) with the non-holomorphic dependence of $F_g^{\rm closed}$. The answer is that the quantity appearing on the right hand is actually the holomorphic limit of $F^{\rm closed}_g$, that is, the limit at which we send $\bar{t}$ to infinity while keeping $t$ finite. In a recent paper, Eynard, Mariño and Orantin \cite{Eynard:2007hf} face up this topic by showing that there is a procedure to obtain non-holomorphic free energies $F^{\rm open}_g(t,\bar{t})$ from the matrix model\footnote{In fact, from any algebraic curve $\Sigma: H(x,y)=0$, without caring whether it is the spectral curve of a matrix model or not.} that satisfy the holomorphic anomaly equations.


In this paper we study the holomorphic anomaly problem concerning eq. (\ref{DVconjecture}) from the point of view of the wave-function interpretation of the topological string partition function. In section 2 we briefly review the real and Kähler polarizations in the quantization of $H^3(M,\real)$. In section 3 we study in detail the process to go both from Kähler to real polarization and the inverse one. The central point of this section is the proof, given recently by Schwarz and Tang \cite{Schwarz:2006br}, that the closed topological string wave-function in real polarization is equal to the holomorphic limit of $Z_{\rm closed}$. In section 4 we re-analyse the results of ref. \cite{Eynard:2007hf} in terms of the $H^3$-quantization formalism. This lets us formulate the Dijkgraaf-Vafa conjecture in a precise background-independent way. Conclusions and comments on the relation to some other topics, like supersymmetric black holes, are given in section 5.   

\section{Wavefuncion interpretation of closed topological strings}

This review section is based on \cite{Witten:1993ed,Gerasimov:2004yx,Verlinde:2004ck,Loran:2005ma,Aganagic:2006wq,Gomez:2006gq}. Let us consider a 7d field theory for a real 3-form $C$ with action
\be
\label{act}
S[C]=\frac{1}{2} \int_{M \times \real} C \wedge d_{7d} C=\int_{M\times\real} \left[\frac{1}{2}\gamma (-\dot{\gamma}+d\omega) + \frac{1}{2} \omega \wedge d\gamma   \right] \wedge dt^\prime 
\ee
where we have the following 6d decomposition
\be
C= \gamma + \omega \wedge d t^\prime
\ee
being $\g$ and $\omega$ real 3 and 2-forms on $M$. For the moment we will consider $M$ to be a compact Calabi-Yau threefold. This is a singular system with conjugate momenta $\pi_\g=-\g/2$ and $\pi_{\omega}=0$. Therefore, the Hamiltonian description is that of a constrained system
with the constraints
\bea
\Phi^{(1)}_\gamma&\equiv &\pi_\gamma + \frac{1}{2}\gamma=0 \label{firstc}\\
\Phi^{(1)}_\omega&\equiv &\pi_\omega=0\\
\Phi^{(2)}_\omega&\equiv &d\gamma=0
\eea
The first two constraints are primary constraints, whereas the last one is a secondary constraint obtained from the second one. Both the second and the third one are first class constraints, and one has to take into account this fact in order to quantize the theory. Thus, the wave functions will not depend on $\omega$ and its dependence on $\g$ will be such that
\be
\label{fisico}
\widehat{d\gamma} | \psi \ra=0
\ee 
On the other hand, eq. (\ref{firstc}) is a set of second class constraints implying that one has to work with Dirac brackets instead of Poisson brackets. From all these constraints one finds that
$H^3(M,\real)$ is the physical phase space of the system.

\subsection{Quantization of {\boldmath$H^3(CY_3,\real)$} in real polarization}

By choosing a symplectic basis $(\a_I,\b^J)$ of $H^3(M)$, with $I,J=0,1,...,h_{2,1}$, one can work with real polarization coordinates
\be
\g=p^I\a_I + q_I \bet^I \in H^3(M,\real)
\ee
From the Dirac brackets one obtains the quantization rule
\be
\label{Heirel}
[q_I,p^J]=i\hbar \d^J_I
\ee
that is, they behave as ordinary coordinate and momentum operators. Under a symplectic transformation they transform as\footnote{Of course, a subgroup of these transformations is the modular group $\Gamma$, but here we would like to stress that one can consider the larger group $Sp(2h_{2,1}+2,\real)$.}
\be
\left( \ba{c}
\tilde{p}\\
\tilde{q}
\ea
\right)=
\left( \ba{cc}
\ensuremath{{\cal{D}}}&\ensuremath{{\cal{C}}}\\
\ensuremath{{\cal{B}}}&\ensuremath{{\cal{A}}}
\ea
\right)
\left( \ba{c}
p\\
q
\ea
\right)
\label{simpletras}
\ee
with $\ensuremath{{\cal{D}}}\ensuremath{{\cal{A}}}-
\ensuremath{{\cal{C}}}\ensuremath{{\cal{B}}}=\imaginario$. This is a canonical transformation, with generating function
\be
S(p,\tilde{p})= -\frac{1}{2}p\ensuremath{{\cal{C}}}^{-1}\ensuremath{{\cal{D}}}p +p\ensuremath{{\cal{C}}}^{-1}\tilde{p} - \frac{1}{2}\tilde{p}\ensuremath{{\cal{A}}}\ensuremath{{\cal{C}}}^{-1}\tilde{p}
\ee
Therefore, wavefunctions on this real polarization $\la \psi | p\ra$ will not be symplectic invariant, but will have this generalized Fourier transformation
\be
\label{genFour}
\la \psi | \tilde{p} \ra = \frac{1}{(2\pi \hbar)^{\frac{h_{2,1}+1}{2}}} \int dp \la \psi | {p} \ra \exp \left[ -\frac{i}{\hbar} S(p,\tilde{p}) \right]
\ee
In the WKB approximation we can write the wavefunction as a series expansion
\be
\la \psi | {p} \ra = \exp \sum_{g=0} \hbar^{g-1} \varphi_g (p)
\ee
Expanding into the leading order saddle point $p_{\rm cl}(\tilde{p})$, which is the solution of
\be
\frac{\pd \varphi_0(p)}{\pd p} -i \frac{\pd S(p,\tilde{p})}{p}=0
\ee
the integral expression (\ref{genFour}) reduces to
\be
\tilde{\varphi_g}(\tilde{p})= \varphi(p_{\rm cl}) + \Gamma_{g}\left[ \Delta^{IJ},\pd_{I_1,...,I_n}\varphi_{r<g}(p_{\rm cl})  \right] \label{symtras}
\ee
where $\Gamma_{g}$ are given by Feynman diagrams \cite{Aganagic:2006wq} with inverse propagator
\be
\Delta_{IJ}=i\frac{\pd^2 \varphi_0}{\pd p^I\pd p^J}(p_{\rm cl}) -(\ensuremath{{\cal{C}}}^{-1}\ensuremath{{\cal{D}}})_{IJ}
\ee
and vertices $\pd_{I_1,...,I_n}\varphi_{r<g}(p_{\rm cl}) $.

\subsection{Quantization of {\boldmath$H^3(CY_3,\real)$} in Kähler polarization}

On the other hand, we can work in a symplectic invariant way by choosing a complex structure on $M$. This induces a polarization on $H^3$, from which we define the Kähler coordinates $\l^{-1}$ and $x^i$, $i=1,...,h_{2,1}$
\be
\g= \l^{-1}\Omega + x^i \De_i \Omega + {\rm cc}
\ee
In these coordinates the commutators coming from the Dirac brackets are
\bea
\label{compalge1}
\left[\l^{-1},\bar{\l}^{-1}\right] &=&-\hbar e^K\\
\left[x^i,\bar{x}^{\bar{j}}\right] &=&\hbar e^K G^{i\bar{j}} \nonumber
\eea
where $K$ is the Kähler potential of the moduli space of complex structures on $M$ and $G^{i\bar{j}}$ is the inverse metric. Notice that $\bar{\l}^{-1}$ and $x^i$ act as annihilation operators. But in order to establish the connection with topological strings it is necessary to work formally with the Hilbert space spanned by the eigenstates $|\l^{-1},x\ra$ of $\hat{\l}^{-1}$ and $\hat{x}^i$
\bea
\label{badco}
| x,\l^{-1} \ra  = \exp \left[ -\frac{1}{\hbar} e^{-K} \hat{\bar{\lambda}}^{-1}{\lambda}^{-1} + \frac{1}{\hbar} e^{-K} x^i\hat{\bar{x}}^{\bar{j}} G_{i\bar{j}}   \right] | 0,0 \ra &\\
\imaginario =\int d\mu_{x,\l^{-1}} \exp \left[ +\frac{1}{\hbar} e^{-K} \bar{\lambda}^{-1}{\lambda}^{-1} - \frac{1}{\hbar} e^{-K} x^i{\bar{x}}^{\bar{j}} G_{i\bar{j}}   \right] | x,\l^{-1} \ra \la \bar{x},\bar{\l}^{-1} |  &\\
\frac{\la \bar{x}^\prime,\bar{\l}^{-1\prime} | x,\l^{-1} \ra }{\la \bar{0},\bar{0} |0,0 \ra }=\exp \left[ -\frac{1}{\hbar} e^{-K} {\bar{\lambda}}^{-1\prime}{\lambda}^{-1} + \frac{1}{\hbar} e^{-K} x^i{\bar{x}}^{\bar{j}\prime} G_{i\bar{j}}   \right] &
\eea
where $d\mu_{x,\l^{-1}}=
|G|^{1/2} \exp\left[-(h_{2,1}+1)K/2\right]
 d^{h}xd^h\bar{x} d\l^{-1}d\bar{\l}^{-1}$.


Another way of describing these states is by using big phase space variables
$\frac{1}{2}x^I=\l^{-1}X^I + x^i \De_i X^I$. That is,
\bea
p^I &=&\R x^I\\
q_I &=&\R \left[ \tau_{IJ}(X) x^J \right]
\eea
Notice that one has to choose a particular symplectic homology basis in order to work with big phase space variables. The quantization rule in these variables is
\be
\left[ x^I,\bar{x}^J \right] = 2\hbar \left[ \I \t(X)  \right]^{-1IJ}
\ee
I will use both notations to denote the same state
\be
\label{igua}
| {x}^I \ra= |x^i,\l^{-1}\ra
\ee
Since
\be
\la p|x\ra= \sqrt{|\I \tau|} \exp \left[ -\frac{i}{2\hbar} p \bar{\tau} p + \frac{1}{\hbar} p \I \tau x-\frac{1}{4\hbar} x\I \tau x  \right]
\ee
the relation between wavefunctions in real and Kähler polarizations is
\be
\la \psi | x\ra = \sqrt{|\I \tau|} \int dp \la \psi | p \ra \exp \left[ -\frac{i}{\hbar} \hat{S}(p,x) \right] \label{realtokahler}
\ee
where
\be
\hat{S}(p,x)= \frac{1}{2} p \bar{\tau} p + i p \I \tau x - \frac{i}{4} x \I \tau x
\ee
is the generating function of the (background dependent) canonical lineal transformation going from real to Kähler polarization.

From the point of view of the real polarization, the eigenstates $|x^i,\l^{-1}\ra_{X,\bar{X}}$ are actually squeezed states $|x^i_{p,q;X,\bar{X}},\l^{-1}_{p,q;X,\bar{X}}\ra_{X,\bar{X}}$ centered around the phase space point $(p,q)$ with width, measuring the quantum resolution, and squeezing parameters given by $\tau_{IJ}(X)$. This is another way to see that these states will change under variations of the base complex structure. It has been shown \cite{Witten:1993ed,Verlinde:2004ck} that the variation of these states is the same as the one of the topological string generating function of correlators as given by the holomorphic anomaly. This suggests to define a state $|\psi_{\rm closed}\ra$ such that its squeezed state representation is equal to the topological string generating function. More precisely
\be
\la \psi_{\rm closed}|\l^{-1},x\ra_{X,\bar{X}}=e^{f_1(X)} \psi_{\rm gen} \left( \sqrt{\hbar}\l,\l x; X,\bar{X} \right) \label{closedstate}
\ee
where $f_1$ is the purely holomorphic part of the genus one free energy. Moreover, it has also been shown that $|\psi_{\rm closed}\ra$ is a physical state of the system \cite{Gerasimov:2004yx} , that is, one that satisfies (\ref{fisico}).

\section{Closed topological string state in real polarization}


In this section we address the problem of computing the wavefunction corresponding to the state $| \psi_{\rm closed} \ra$ in real polarization. This computation was done in an elegant way by Schwarz and Tang \cite{Schwarz:2006br} by introducing, as an auxiliary tool, a hybrid polarization, which mixes real and Kähler bases. We classify and describe the four possibilities of doing this mix in the next subsection. In this paper we use the name ``holomorphic'' or ``anti-holomorphic'' for these hybrid polarizations, depending on whether its background dependence is holomorphic or anti-holomorphic.

\subsection{(Anti-)Holomorphic polarizations}

\subsubsection{$(\Omega,\beta)$-holomorphic polarization}
The polarization we are interested in is \cite{Schwarz:2006br}
\be
\label{holopo}
\gamma = \frac{1}{2} x^I_{\rm hol} \pd_{I} \Omega + q_{{\rm hol} I} \b^I
\ee
It is straightforward to obtain that
\bea
x^I_{\rm hol} &=&2p^I \label{xesp}\\
q_{{\rm hol} I} &=& -i \I \tau_{IJ} \bar{x}^J
\eea
and, therefore
\be
\left[ q_{{\rm hol} I}, x^J_{\rm hol}  \right]= 2i\hbar \delta^J_I
\ee
From (\ref{xesp}) we trivially have that $|p\ra$ are the eigenstates of $\hat{x}^I_{\rm hol}$. Therefore wavefunctions in the $x_{\rm hol}$-representation are nothing but $\la \psi | p\ra$. Nevertheless, we shall use $|x_{\rm hol}\ra$ since they have a different natural normalization factor. By writing
\be
|x_{\rm hol}\ra = \exp \left[ \frac{i}{2\hbar} x_{\rm hol} \hat{q}_{\rm hol}\right] |x_{\rm hol}=0\ra
\ee
we find
\be
|x_{\rm hol}\ra=\exp \left[  -\frac{i}{2\hbar} p \tau p \right] | p\ra \label{holostates}
\ee
The base point dependence is
\bea
\frac{\pd}{\pd X^J} |x_{\rm hol}\ra &=& -\frac{i}{8\hbar} C_{IJK} x^I_{\rm hol} x^K_{\rm hol}  |x_{\rm hol}\ra\\
\frac{\pd}{\pd \bar{X}^J} |x_{\rm hol}\ra &=& 0
\eea

We can also introduce the $(\l,x^i)$ notation
\be
\g= \l^{-1}_{\rm hol} \Omega + x^i_{\rm hol} \De_i \Omega +q_{{\rm hol}I}\b^I
\ee
where
\be
\frac{x_{\rm hol}^I}{2}= \l^{-1}_{\rm hol} X^I + x^i_{\rm hol} \De_i X^I
\ee
The base point dependence is then given by
\bea
\frac{\pd}{\pd t^i} |\l^{-1}_{\rm hol},x^i_{\rm hol}\ra &=& \left[ \l^{-1}_{\rm hol} \frac{\pd}{\pd x^i_{\rm hol}} - \frac{1}{2\hbar} C_{ijk} x^j_{\rm hol} x^k_{\rm hol}  \right] |\l^{-1}_{\rm hol},x^i_{\rm hol}\ra\\
\frac{\pd}{\pd \bar{t}^{\bar{i}}} |\l^{-1}_{\rm hol},x^i_{\rm hol}\ra &=& 0
\eea

\subsubsection{Other (anti-)holomorphic polarizations}

We can also find in the literature the antiholomorphic polarization \cite{Gunaydin:2006bz}
\be
\label{piolin}
\gamma = \frac{1}{2} y^I \bar{\pd}_{I} \bar{\Omega} + s_{I} \b^I
\ee
for which
\bea
y^I &=&2p^I \label{yesp}\\
s_{I} &=& i \I \tau_{IJ} {x}^J
\eea
and, therefore
\be
\left[ s_{I}, y^J  \right]= 2i\hbar \delta^J_I
\ee
Now $|p\ra$ are the eigenstates of $\hat{y}^I$, and the eigenstates of $\hat{s}_I$ are $|x\ra$, so this formalism contains both the real and the Kähler polarizations. With the natural normalization factor we have
\be
|s\ra = \frac{1}{\sqrt{|\I \t|}} \exp \left[ \frac{1}{4\hbar} x \I \t x\right] |x\ra \label{piowf}
\ee
The base point dependence is
\bea
\frac{\pd}{\pd {X}^J} |s\ra &=& 0\\
\frac{\pd}{\pd \bar{X}^J} |s\ra &=& -\frac{i}{2\hbar} \bar{C}_{IJK} \hat{p}^I\hat{p}^K  |s\ra
\eea
From (\ref{piowf}) and (\ref{closedstate}) one can find that
\be
\la \psi_{\rm closed} | s \ra_{\bar{X}}= \exp \left[ \frac{i}{4\hbar} x\bar{\t} x -\bar{f}_1(\bar{X}) +\sum_{g=0} \hbar^{g-1} F^{\rm closed} (\frac{x}{2},\bar{X})  \right]
\ee
Therefore, the reason why $\la \psi_{\rm closed} | s \ra_{\bar{X}}$ has only an antiholomorphic background dependence is because the holomorphic dependence has been absorbed into the wavefunction dependence.

The other two possibilities are
\be
\label{posd}
\gamma = \frac{1}{2} w^I \bar{\pd}_{I} \bar{\Omega} + p^{I}_{\rm hol} \a_I
\ee
for which
\bea
w^I &=&2\bar{\t}^{-1 JI} q_I \label{wesq}\\
p_{\rm hol} &=& -i \bar{\t}^{-1}(\I \tau)^{-1} {x}
\eea
and
\be
\label{posc}
\gamma = \frac{1}{2} u^I {\pd}_{I} {\Omega} + r^I \a_I
\ee
for which
\bea
u &=&2\t^{-1}q \label{uesq}\\
r &=& i \t^{-1}(\I \tau)^{-1} \bar{x}
\eea

\subsection{Loss of background dependence: the $\bar{z}\to\infty$ limit}

The relation between holomorphic (\ref{holopo}) and Kähler polarization bases is given by
\be
\bar{\pd}_I \bar{\Omega}= \pd_{I} \Omega - 2i \I \tau_{IJ} \b^J 
\ee
Thus, both bases will be the same in the limit where
\be
\frac{i}{2} \left[(\I \tau)^{-1}\right]^{IJ}\pd_J \Omega
\ee
is small. By doing the wedge product with the elements of the symplectic basis $(\b^I,\a_J)$ one obtains the conditions
\bea
 &\frac{i}{2}(\I \tau)^{-1} &\simeq 0\\
&\frac{i}{2}(\I \tau)^{-1}\tau &\simeq 0
\eea
that is,
\be
\t_{IJ} + \bar{\t}_{IJ} \simeq -\tau_{IJ} + \bar{\t}_{IJ} \to \pm \infty \label{condi}
\ee
Of course, this limit cannot be satisfied if one keeps $\bar{t}$ to be the complex conjugate of $t$. The way to satisfy (\ref{condi}) is by sending
\bea
z & \to & s \\
\bar{z} & \to & \nu \bar{s} 
\eea
with $\nu \to \infty$ and $s$ a complex constant. $(z,\bar{z})$ are the coordinates on the complex structure moduli space that give the Kähler parameters. In other words, $z$ is kept fixed whereas $\bar{z}$ is sent deep inside the Kähler cone. In this limit
\be
\bar{\pd}_{I} \bar{\Omega} \to -2i \I\t_{IJ} \b^J
\ee
and the Kähler operators go to
\bea
x^I &\to & x^I_{\rm hol}\\
-i \I\t_{IJ}\bar{x}^J & \to & q_{{\rm hol}I}
\eea
States $|x^I \ra$ and $|x^I_{\rm hol} \ra$ will be proportional. With the normalizations we have chosen the proportionality constant is indeed one
\be
|x^I\ra_{X,\bar{X}_\infty}=|x^I_{\rm hol}\ra_{X}  \label{limitstates}
\ee
From (\ref{limitstates}) and (\ref{closedstate}) we have
\bea
&& \la\psi_{\rm closed} | \l^{-1}_{\rm hol}, x^i_{\rm hol} \ra_{X}  = \label{ecu1} \\
&& =\exp \left[ f_1(X) + \sum_{g=0} (\l_{\rm hol}\sqrt{\hbar})^{2g-2} \sum_{n=0} \frac{1}{n!} C^{g}_{i_1...i_n}(X,\bar{X}_\infty) (\l_{\rm hol} x_{\rm hol}^{i_1})...(\l_{\rm hol} x_{\rm hol}^{i_n}) \right]\nonumber 
\eea
Notice that the last expression does not contain the genus 0 free energy. This is due to the selection rules of the topological string correlators. Since in the holomorphic limit $\pd_{i}K\propto \frac{1}{\nu}\to 0$, the relation between Kähler and big phase space variables is simpler
\bea
\frac{1}{2}x^0_{\rm hol} &=& \l^{-1}_{\rm hol} X^0\\
\frac{1}{2}x^{I=i}_{\rm hol} & = & X^0 \left( \l^{-1}_{\rm hol} t^i + x^i\right)
\eea
where we have chosen coordinates $t^i=\frac{X^i}{X^0}$. Eq. (\ref{ecu1}) becomes
\bea
\la\psi_{\rm closed} | \l^{-1}_{\rm hol}, x^i_{\rm hol} \ra_{X} & = &
 \exp \left[ f_1(X) + \sum_{g=0} (\l_{\rm hol}\sqrt{\hbar})^{2g-2} F_{g}^{\rm closed} \left(\frac{x_{\rm hol}}{2},\bar{X}_{\infty} \right) - \right. \nonumber\\
 &  & \left. -(\l_{\rm hol}\sqrt{\hbar}) ^{-2} F_0^{\rm closed} (X) -(\l_{\rm hol}\sqrt{\hbar})^{-2} (\l_{\rm hol}x^i_{\rm hol})  \pd_i F_0^{\rm closed} (X)- \right.\nonumber\\
& & \left. -\frac{1}{2}(\l_{\rm hol}\sqrt{\hbar}) ^{-2} (\l_{\rm hol}x^i_{\rm hol}) (\l_{\rm hol}x^j_{\rm hol})  \pd_i\pd_j F_0^{\rm closed} (X) \right] \label{ecu2}
\eea
On the other hand we have
\bea
-\frac{i}{2\hbar} \frac{x^I_{\rm hol}}{2} \t_{IJ} \frac{x^I_{\rm hol}}{2} & = &
-(\l_{\rm hol}\sqrt{\hbar}) ^{-2} F_0^{\rm closed} (X) -(\l_{\rm hol}\sqrt{\hbar})^{-2} (\l_{\rm hol}x^i_{\rm hol})  \pd_i F_0^{\rm closed} (X)- \nonumber\\
& &  -\frac{1}{2}(\l_{\rm hol}\sqrt{\hbar}) ^{-2} (\l_{\rm hol}x^i_{\rm hol}) (\l_{\rm hol}x^j_{\rm hol})  \pd_i\pd_j F_0^{\rm closed} (X)  \label{ecu3}
\eea
Combining eq. (\ref{holostates}), (\ref{ecu2}) and (\ref{ecu3}) we obtain a simple expression for the closed topological string state in real polarization
\be
\la \psi_{\rm closed} | p \ra = \exp \left[  \sum_{g=0} \hbar^{g-1} F_g^{\rm closed}(p,\bar{X}_{\infty})   \right]
\ee
In conclusion, we can see that, in the process in order to go from Kähler to real polarization, the background dependence is lost by
\begin{itemize}
\item sending the antiholomorphic dependence to infinity and by
\item treating the holomorphic dependence as the functional dependence $\psi (p)$ of the wavefunction.
\end{itemize}

\subsection{Loss of symplectic dependence}

One way to see what happens in the inverse process, i.e. to go from real to Kähler polarization, is to use the Feynmann diagrams of ref. \cite{Aganagic:2006wq}. Let us consider for simplicity eq. (\ref{realtokahler}) in the particular case $x^I=2\l^{-1}X^I$. This is the particular background point at which the ``attractor equations''
\bea
p^I &=&\R \left[  2\l^{-1}X^I \right] \label{at1}\\
q_I &=&\R \left[ 2\l^{-1}\tau_{IJ} X^J \right]  \nonumber
\eea
hold. The pair $(p,q)$ is the phase space point at which the squeezed states $| x\ra_{X,\bar{X}}$ are centered. Therefore we are studying
\be
\la \psi | x=2\l^{-1}X \ra_{X,\bar{X}} \ra = \exp \left[ f_1(X) + \sum_{g=2} (\l\sqrt{\hbar})^{2g-2} F_g^{\rm closed} (X,\bar{X}) \right]
\ee
in terms of its real polarization counterpart. Expanding the integral of eq. (\ref{realtokahler}) into the leading order saddle point
\be
p_{\rm cl} =\l^{-1}X
\ee
one finds
\bea
&& \la \psi | x=2\l^{-1}X \ra_{X,\bar{X}} \ra = \label{ecu4} \\
&& = \exp \left[ f_1(X) + \sum_{g=2} (\l\sqrt{\hbar})^{2g-2} \left( F_g^{\rm closed} (X,\bar{X}_{\infty}) + \Gamma_g ((-2i\I \tau)^{-1}, \pd_{I_1}...\pd_{In} F_{r<g}^{\rm closed}(X,\bar{X}_\infty) ) \right) \right]   \nonumber
\eea
where $\Gamma_g$ are the same Feynman diagrams that appear in eq. (\ref{symtras}), but with a different propagator
\be
\label{difprop}
\breve{\Delta}_{IK}(X)=-2i\I\t(X)
\ee
Notice that there is not $\hbar^{-1}$ term into eq. (\ref{ecu4}). This is because the leading order saddle poit evaluation of the integral (\ref{realtokahler}) is equal to 1. In addition, the 1-loop term
\be
\Gamma_1= -\frac{1}{2} \log |\I \t|
\ee
cancels with the $|\I\tau|$ that is in front of the integral (\ref{realtokahler}). From eq. (\ref{ecu4}) the conclusion is that the non-holomorphic dependence of $\la \psi_{\rm closed}|X\ra_{X,\bar{X}}$ comes enterely from the propagators (\ref{difprop}) of Feynman diagrams. On the other hand, $F_g^{\rm closed}(X,\bar{X}_\infty)$ transforms in a specific way under symplectic transformations (\ref{symtras}), whereas $\la \psi_{\rm closed}|X\ra_{X,\bar{X}}$ is clearly, up to a normalization constant, symplectic invariant. This is due to the fact that the propagator transforms as
\be
(-2i \I \t)^{-1} \to (\Ce \t + \De)^I_K (-2i \I \t)^{-1KL}(\Ce \t + \De)^J_L -(\Ce\tau + \De)^J_L\Ce^{IL} 
\ee
in such a way that this quasi-modular transformation cancels the transformations of $F_g^{\rm closed}(X,\bar{X}_\infty)$.

\section{Matrix model partition function as a real polarization wavefunction}

Everything that has been said until now for the quantization of $H^3(M,\real)$ can be extrapolated, up to some subtleties, to the case where $M$ is a local Calabi-Yau. We will center on the concrete class of local CY backgrounds of ref. \cite{Dijkgraaf:2002fc}
\bea
uv=H(x,y); & & H(x,y)= y^2 - (W^\prime(x))^2 + f(x)
\eea
where $f(x)$ is a polynomial of degree $d-1$. Their coefficients parametrize the complex structure deformations over the singular manifold. We call this deformed manifold $M_{\rm def}$. Its geometry can be seen as a $C^*$ fibration over the $xy$-plane. 3-cycles on $M_{\rm def}$ descend to 1-cycles on the hyperelliptic surface $\Sigma: H(x,y)=0$, and periods of the holomorphic 3-form $\Omega$ on $M_{\rm def}$ descend to the periods of a meromorphic 1-form on $\Sigma$. To simplify notation we will use the same letters for 3-cycles and 1-cycles, and we will also call the meromorphic 1-form $\Omega$. 

For these manifolds, we can consider $2d-2$ compact 1-cycles $(A^i,B_j)$, with $i=1,2,...,d-1$, forming a symplectic basis.  But, in addition, there are cycles $\hat{A}$ whose homology dual cycles $\hat{B}$ are non-compact. This is the reason why, whereas the quantities
\be
X^i=\int_{A^i} \Omega
\ee
are complex structure moduli giving rise together with
\be
F_i=\int_{B_i} \Omega
\ee
to the usual rigid special geometry relations, the quantities
\be
\hat{X}=\int_{\hat{A}}\Omega
\ee
are considered as parameters on the model, not moduli. The useful basis for us is the one of ref. \cite{Bilal:2005hk}, at which there is only one non-compact cycle $\hat{B}$.

For the moment, we are going to consider the quantization of the component of $\gamma$ that is a linear combination of the  forms $(\b^i,\a_j)$, which are the Poincaré dual of $(A^i,B_j)$. Everything works as explained in previous sections, but instead of having $I=0,1,...,h_{2,1}$, we have $i=1,...,d-1$. The phase space coordinates $(p^i,q_j)$ are promoted to operators, whereas $\hat{p}$ is treated as a given parameter. In particular, eq. (\ref{symtras}) still works, but refers to elements of the symplectic group $Sp(2d-2,\real)$.

On the other hand, in ref. \cite{Eynard:2007kz} it is shown how to construct recursively a set of scalar functions $\underline{F_g^{\rm H}}(X^i)$ from the curve $\Sigma$. In the case we are considering, they are precisely the free energies of the matrix model (\ref{mamo}), whose classical spectral curve is $\Sigma$. The procedure of ref. \cite{Eynard:2007kz} uses a modified Bergmann kernel to compute modified functions $F_g^{\rm H}$. This modified Bergmann kernel depends on a symmetric matrix $\kappa$ in such a way that
\be
\underline{F_g^{\rm H}}=F_g^{\rm H} \arrowvert_{\kappa = 0}
\label{cuatrocinco}
\ee
Ref. \cite{Eynard:2007hf} uses the variations $\frac{\pd F_g^{\rm H}}{\pd \kappa}$, computed in \cite{Eynard:2007kz}, for the particular case
\be
\kappa^{ij}=\left(-2i \I \t  \right)^{-1ij}=\breve{\Delta}^{ij}
\ee
and shows that
\be
\underline{F^{\rm H}_g}({X^i})\to {F_g^{\rm H}}(X^i,\bar{X}^{\bar{i}})=\underline{F_g^{\rm H}}(X^i) + \Gamma_{g}\left[ \kappa^{ij},\pd_{I_1,...,I_m}\underline{F_{r<g}^{\rm H}}(X^i)  \right] 
\ee
or, analogously, that the new $F_g^{\rm H}$ verify the holomorphic anomaly equations. It was also known from \cite{Eynard:2007kz} that with this choice of $\kappa$ the new $F_g^{\rm H}$ are modular invariant because the Bergmann kernel is. From the point of view of the quantization of $H^3(M,\real)$ this is nothing but the transformation (\ref{ecu4}). Notice that the choice of $\kappa$ is the one corresponding to the canonical change of variables going from real to Kähler polarization.

From the previous discussion, it is clear that, by doing the same analysis for the case we consider a general modular transformation (\ref{simpletras}), the unmodified quantities $\underline{F_g^{\rm H}}$ change as
\be
\underline{F_g^{\rm H}}(X^i) \to \underline{F_g^{\rm H}}(X^i_{\rm cl}) + \Gamma_{g}\left[ \Delta^{ij},\pd_{I_1,...,I_m}\underline{F_{r<g}^{ \rm H}}(X^i_{\rm cl})  \right] 
\ee
We can see it by noticing that the modular transformed $\underline{F_g^{\rm H}}$ are equal to $\underline{F_g^{\rm H}}\arrowvert_{\kappa=\Delta}$. Thus, the quantities $\underline{F_g^{\rm H}}$ transform  in the same way as $\varphi_g$ (see eq. (\ref{symtras})). We saw in section 2 that they are the only conditions the functions $\varphi_g$ must satisfy in order to represent a background independent and symplectic-modular invariant state $|\psi\ra$ belonging to the naive Hilbert space of the quantization of $H^3(M,\real)$. For this reason, we propose to associate to any given algebraic curve $H(x,y)=0$ a state $|\psi_{\rm H}\ra$ such that
\be
\label{defin}
\la \psi_{\rm H} | {p} \ra = \exp \sum_{g=0} \hbar^{g-1} \underline{F^{\rm H}_g} (p)
\ee
is its momentum representation. In the case we are considering, where $\Sigma$ is the spectral curve of a matrix model, we denote this state by $|\psi_{\rm open}\ra$.

In addition, the conclusion of \cite{Eynard:2007hf} is that the quantities $F_g^{\rm H}(X,\bar{X})$ are equal to $F_g^{\rm closed}(X,\bar{X})$ up to a holomorphic modular invariant quantity. Therefore, in order to prove the Dijkgraaf-Vafa conjecture the thing that remains to show is that this holomorphic modular invariant quantities are equal to zero at all genera. This should be done, at least in principle, by impossing the appropiate boundary behaviour at the conifold point of the complex structure moduli space. Now, with the definition (\ref{defin}) and the results of section 3, this is the same as saying that
\be
\label{conjetura}
|\psi_{\rm H}\ra=|\psi_{\rm closed}\ra
\ee
This is not a crazy stament because both states are defined as a topological property of the surface $H(x,y)=0$. Notice however, that, although they are topological invariants of $M_{\rm def}$, their origin is much different:
\begin{itemize}
\item $|\psi_{\rm closed}\ra$ comes from closed topological strings on $M_{\rm def}$ with a specific complex structure $(\Omega,\bar{\Omega})$. This is the reason why we can say that the natural polarization associated with closed strings on $M_{\rm def}$ is $|\l^{-1},x^i\ra_{\Omega,\bar{\Omega}}$.
\item $|\psi_{\rm H}\ra$ comes from the invariant functions $\underline{F_g^{\rm H}}(p;(A,B))$, which can be obtained from $M_{\rm def}$ by choosing a symplectic basis $(A,B)$. These functions do not depend on the complex structure of $M_{\rm def}$.  This is the reason why we can say that the natural polarization associated with these invariants is $|p\ra_{(A,B)}$.
\end{itemize}
Notice also that, in particular, the conjecture (\ref{conjetura}) implies that $|\psi_H\ra$ is actually a physical state, i.e. one that satisfies the condition (\ref{fisico}).

On the other hand, in the open string side we do not have the freedom to choose a symplectic structure. In this context, this can be understood from the fact that free energies $F^{\rm open}$ are equal to $\underline{F^{\rm H}}$ at a fixed symplectic basis where the $A$-periods are proportional to the filling fractions
\be
\int_{A_{\rm o}^i} \Omega\propto \nu^i
\ee
Thus, the natural polarization associated with open strings on $M_{\rm res}$ is $|p\ra_{(A_{\rm o},B_{\rm o})}$, and the precise definition of $|\psi_{\rm open}\ra$ is
\be
\label{deopen}
\la \psi_{\rm open} | {p} \ra_{(A_{\rm o},B_{\rm o})} = \exp \sum_{g=0} \hbar^{g-1} F^{\rm open}_g (p)
\ee

\section{Conclusions and discussion}

We have seen that we can associate, both to the open string background and to the closed one, states in the quantization of $H^3(M_{\rm def},\real)$ in such a way that the Dijkgraaf-Vafa conjecture reads
\be
|\psi_{\rm open}\ra =|\psi_{\rm closed}\ra 
\ee
If the conjecture is true, we see that open and closed string amplitudes are nothing but different representations of the same background independent state. This is the reason why the geometric transition process that goes from open to close string backgrounds is, from this point of view, a change from real to Kähler polarization. On the left-hand side (open strings) wavefunctions are holomorphic, but change under modular transformations. On the right-hand side (closed strings) they have a non-holomorphic dependence, but they are modular invariant. In order to see that this is the natural way to look at this brane/flux geometric transition, we have pointed out that
\begin{itemize}
\item In the closed-string side of the duality we have a target space geometry with background complex structure $(\Omega,\bar{\Omega})$ but without any privileged symplectic basis. The symplectic basis is introduced only through the definition of the periods $X^i$.
\item On the other hand, in the open string side, the resolved geometry does not have the complex structure moduli $X^i$, which have been replaced by branes. We have lost that background dependence. Nevertheless, at this open string side, the information about $X^i$ is encoded into the filling fractions, so there is a privileged symplectic basis given by the numbers of branes at the different $\complejo P^1$s of the geometry: we are in a real polarization description.
\end{itemize}


In addition, we would like to point out that proposals (\ref{defin}) and (\ref{conjetura}) can be extended naturally to include the non-compact sector, in such a way that $i=1,2,...,d,...$ by considering the non-compact cycle as the limit of a compact one, and by considering also the dependence of the matrix model free energies on the 't Hooft parameter associated with the total size of the matrix\footnote{In fact, in ref. \cite{Bilal:2005hk} it is shown how to work with cut-off dependent quantities associated with the non-compact cycle $\hat{B}$, and how the special geometry relations are modified when one includes $\hat{B}$ in the analysis.}. Nevertheless, it would be pleasant to formalize the whole analysis by working directly with local CY background without taking any limit. 

We would also like to indicate that all the analysis that was done in section 4 concerning formulas from (\ref{cuatrocinco}) to (\ref{conjetura}) can be extrapolated to any of the algebraic curves considered in \cite{Eynard:2007kz} and, in particular, to the backgrounds of ref. \cite{Marino:2006hs}. In fact the latter backgrounds can also be associated with some limit of the geometric transitions of ref. \cite{Aganagic:2002qg}\footnote{I would like to thank M. Mari\~no for pointing this fact out to me.}. Thus it would be very interesting to extend the present work to include these more complicated open/closed string dualities.

We expect this new way of looking at the geometric transitions to give new insight into the study of supersymmetric black holes in string theory. Macroscopic entropy of the so called Calabi-Yau black holes is related to closed topological string free energies \cite{Ooguri:2004zv,LopesCardoso:2006bg}
 and, therefore, to $|\psi_{\rm closed}\ra$. In fact, in ref. \cite{Gomez:2006gq} is shown that this macroscopic entropy is related to the mixed Husimi-antiHusimi quantum distribution function associated with $|\psi_{\rm closed}\ra$. In this formalism the attractor equations (\ref{at1}) going from Kähler to real polarization play a special role. The pairs $(p,q)$ are, in this case, the charges of the black hole. Therefore they are integer variables. This fact, although usually ignored in the literature about the quantization of $H^3$, is naturally encoded into the matrix model formalism: the quantities $p^i$ represent the number of matrix eigenvalues located at the critical points of the potential $W$. The fact that $p^i$ are integer is also included into the quantization of the curve $H(x,y)=0$ through the relation
\be
\label{quansigma}
\left[ x,iy \right]=i\hbar
\ee
More precisely, $p^i \in \entero$ is the Bohr-Sommerfeld quantization rule associated with the closed phase space curve surrounding the critical point where the eigenvalues are located \cite{Gomez:2005qp}. It is also known that the relation (\ref{quansigma}) is responsible for the wavefunction behavior of the open topological string partition function associated with non-compact branes \cite{Aganagic:2003qj,Aganagic:2005dh}. However in this work we have studied only the case of compact branes, for which the wavefunction behavior is given by (\ref{Heirel}). It would be interesting to study the interplay between both quantizations.

On the other hand, it is precisely the real polarization description the one that is related to Gopakumar-Vafa invariants and that appears in the recent microscopic derivations \cite{Gaiotto:2006ns,Beasley:2006us,deBoer:2006vg,Denef:2007vg} of the Ooguri-Strominger-Vafa conjecture \cite{Ooguri:2004zv}. In these derivations the quantum corrected entropy appears as the Wigner function associated with $|\psi_{\rm closed}\ra$. The usual case considered in the literature is the one where the complex structure attractor point is located in the region deep inside the Kähler cone $(X,\bar{X})\simeq (X_{\infty},\bar{X}_{\infty})$. From conclusions of section 3, it is clear that at this region one is not able to distinguish between real and Kähler polarizations and, in fact, it is shown in ref. \cite{Gomez:2006gq} that in this region the macroscopic entropy of the black hole does not differ significantly from a Wigner function. But, clearly, if one wants to work with black holes that are outside the region $(X,\bar{X})\simeq (X_{\infty},\bar{X}_{\infty})$ one has to take into account the change of polarization needed to compare microscopic and macroscopic entropies.


\acknowledgments
I wish to thank C. G\'omez for reading the manuscript, for making many comments and suggestions and for his ongoing guidance. I also wish to thank M.~Mari\~no, S.~Montero, K.~Landsteiner and J.~Bellor\' \i n for reading the manuscript and making many comments and suggestions. I would also like to thank S.~Cid for her kind encouragement,
support and affection.
This work is supported by Universidad Autónoma de Madrid.


\bibliographystyle{JHEP-2}

\end{document}